# Delamination in Large REBCO Coils

Jun Lu, Yan Xin, Iain Dixon, *Senior Member, IEEE,* Brent Jarvis, Kwangmin Kim, Yu Suetomi, Jeremy Levitan, Jan Jaroszynski, and Hongyu Bai

***Abstract*— REBCO coated conductors have been used in ultrahigh field magnets for condensed matter physics research, nuclear fusion, particle accelerators, and NMR applications. It is well known, however, that they are not mechanically strong against delamination due to their intrinsic layered structures. As a result, the delamination issue has become one of the major design challenges of REBCO magnet coils.**

**As a part of the development of the 40 T all-superconducting magnet at the National High Magnetic Field Laboratory, USA, large-scale coils wound by two-in-hand REBCO conductor with turn-to-turn insulation was designed and tested at 4.2 K in an 11.4 T background magnetic field. We found REBCO degradations within the windings and at the crossover joints during the tests. The postmortem inspection found that the degradation was caused by delamination. The delamination appears on the surface of the conductor as bubbles which were subsequently studied by microscopy. Additionally, a similar bubble was found on a short sample by torque magnetometry. This supports the inference that the delamination bubbles found in coils are due to electromagnetic stress. In this paper, we present these findings. The implication of these findings for future magnet designs and the method of mitigation will be discussed.**

## I. INTRODUCTION

REBCO coated conductors have been used in ultrahigh field magnets for condensed matter physics research, nuclear fusion, particle accelerators, and NMR applications. It is well known, however, that they are not mechanically strong against delamination due to their intrinsic layered structures. As a result, the delamination issue has become one of the major design challenges of REBCO magnet coils.

Severe degradation due to delamination was discovered in epoxy impregnated REBCO coils [1], which stimulated active measurement of REBCO delamination strength by many research groups worldwide [2] – [5].

Data from the anvil test suggest that delamination strength is greater than 10 MPa for most commercial REBCO tapes [2]. Recently, however, experimental evidence showed that delamination can be caused by electromagnetic stress from applied transport current [6], [7] or the induced magnetization current (screening current) [8] and the delamination strength was as low as 3 - 5 MPa. This raises concern for high field magnet design because such a delamination strength is comparable with the electromagnetic stress in REBCO magnets operating above 30 T.

In our previous work, delamination was found in a metal-insulation coil [9]. In this paper, we present our observation of delamination in insulated and wax impregnated large coils. In addition, similar delamination by screening current was demonstrated in our short sample tests by torque magnetometry. We will discuss the implication of these findings for future magnet designs and the method of mitigation.

## II. EXPERIMENT

REBCO conductor used in this work is SuperPower SCS4050-AP grown by IBAD/ MOCVD on 50 µm thick Hastelloy C-276 substrate. It is 4 mm wide tape with a nominal 1.5 µm silver layer and an electroplated copper stabilizer, 20 µm on each side.

Double-pancake modules were wound by the two-in-hand REBCO conductor with a 316L stainless steel and a copper co-wind tape. The so-called large-scale coil (LSC) consisting of 22 modules is a prototype of one of REBCO insert coils for the 40 T all-superconducting magnet project under development at the National High Magnetic Field Laboratory, USA (NHMFL). The coil parameters relevant to this study are listed in Table 1. The coil was insulated by sol-gel coating on stainless steel co-wind and impregnated with paraffin.

TABLE I LSC COIL PARAMETERS

| ID | Value |
|---|---|
| Conductor | Two-in-hand REBCO<br>Both REBCO facing inside |
| Inner diameter | 262.8 mm |
| Outer diameter | 293 - 312 mm |
| Number of modules | 22 |
| Overband | Multi-turns of 316L SS tape |
| Insulation | Sol-gel coating |
| Impregnation | Paraffin |
| Design operating current | 645 A |
| Maximum total field | 16.4 T |

Two tests, named LSC1-1 and LSC1-2, were performed at 4.2 K and in 11.4 T background field in the superconducting

This paragraph of the first footnote will contain the date on which you submitted your paper for review, which is populated by IEEE. This work was performed at the National High Magnetic Field Laboratory, which is supported by National Science Foundation Cooperative Agreement No. DMR-1644779, DMR-1839796, DMR- 2131790, and the State of Florida. *(Corresponding author: Jun Lu*. Email:junlu@magnet.fsu.edu).

Jun Lu, Yan Xin, Iain Dixon, Brent Jarvis, Kwangmin Kim, Yu Suetomi, Jeremy Levitan, Jan Jaroszynski, and Hongyu Bai are with the National High Magnetic Field Laboratory, Tallahassee, FL 32310, USA.

outsert of the NHMFL 45T hybrid magnet [10].

To investigate delamination in short REBCO samples under similar electromagnetic conditions as in LSC coils, a torque magnetometer probe was used [11]. Fig. 1(a) is a schematic of the torque measurement probe. The sample is rotated by pulling the fishing line while measuring the pulling force with a load cell. As illustrated in Fig. 1(b), the sample stage has an unsupported area which allows for delamination. The induced screening current results in electromagnetic tensile/compressive stresses in some areas. Due to the strong anisotropy of the screening current, the thickness-wise tensile stress is at maximum when the sample plane is parallel to magnetic field. The sample was cut from the same spool of conductor as that in the previous study [8]. Copper, silver and REBCO layers were etched by nitric acid from 13 mm long samples except for a ~5 mm long area. The surrounding edge copper in the 5 mm area was also etched for some samples which we call 'island' samples. More details of this technique can be found in [8].

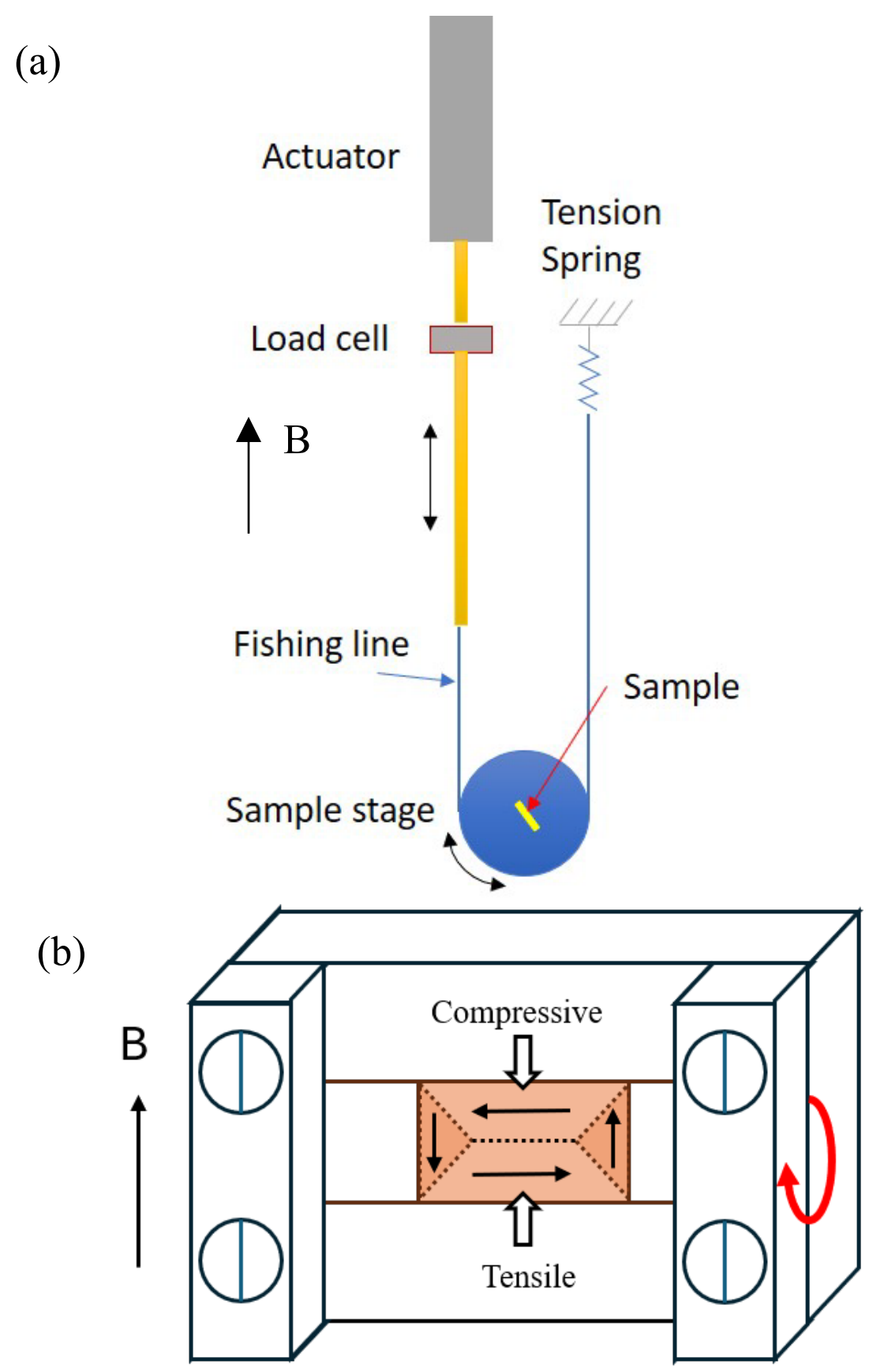

**Fig. 1.** (a) Schematic of a torque magnetometer probe. (b) A sample on the sample stage where directions of screening current and electromagnetic stresses are indicated.

## III. Results

### *A. Delamination in LSC1-1*

The LSC1-1, tested in 2025, showed resistive voltage in some modules at 620 A, short of the designed operating current of 645 A. Those modules were unwound and inspected visually as well as by the reel-to-reel magnetization measurement at 77 K using Tapestar™.

Visual inspection of the 5th module from the top found bubbles on the surface at multiple locations. These bubbles, as shown in Fig. 2(a), seem to be at random locations throughout the module and linked to some levels of $I_c$ degradation at the same location in Tapestar™ data. A 77 K transport $I_c$ measurement of the section in Fig. 2(a) showed about 50% degradation. To investigate inside the bubbles, the edges of this sample were trimmed, and copper/REBCO layer was carefully peeled. The result is shown in Fig. 2(b). Both Cu side and substrate side of the conductor were covered by residual REBCO with no indication of the locations of the bubbles. Furthermore, the copper layer of another sample with surface bubbles was partially removed by the copper etchant APS-100 at room temperature for 20 minutes. After the etching, as shown in Fig. 2(c), the bubbles became a wrinkled band under the silver layer, a clear indication of delamination.

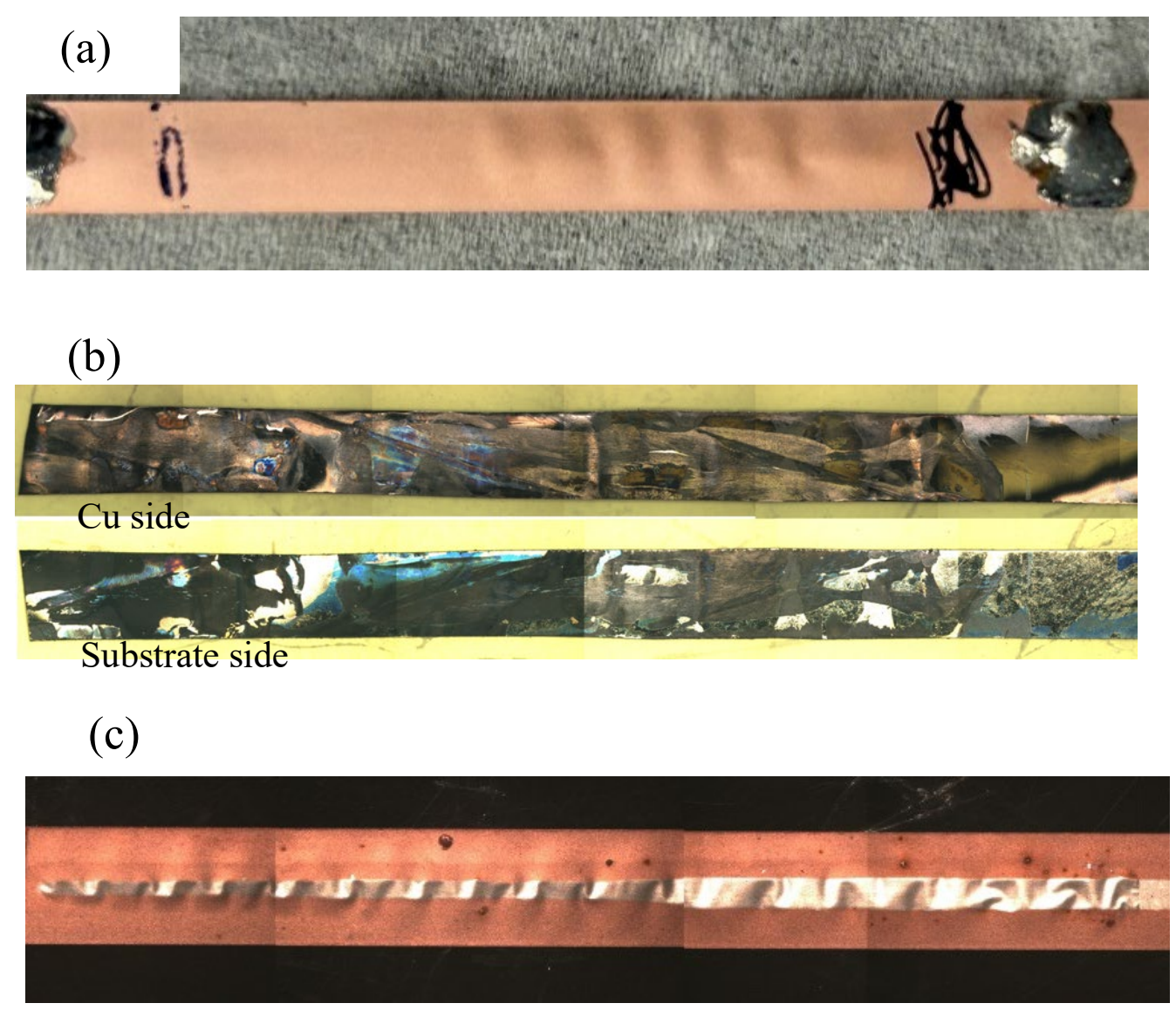

**Fig. 2.** Bubbles observed in module 5 of LSC1-1. (a) five bubbles in a row. (b) after manually delaminating the sample in (a). (c) after copper was partially removed chemically from another sample, the delamination under the bubbles is evident.

A cross-section of the delaminated region was cut by focused ion beam and imaged by a scanning electron microscope (Helios G4 FIB/SEM). The dark line within the REBCO layer in Fig. 3(a) is the delamination crack. In addition, a transmission electron microscope (JEOL JEM-ARM200cF, TEM) was used to examine the perimeter of a bubble. Fig. 3(b) clearly shows a crack propagating from within the REBCO layer and ends at the REBCO/silver interface.

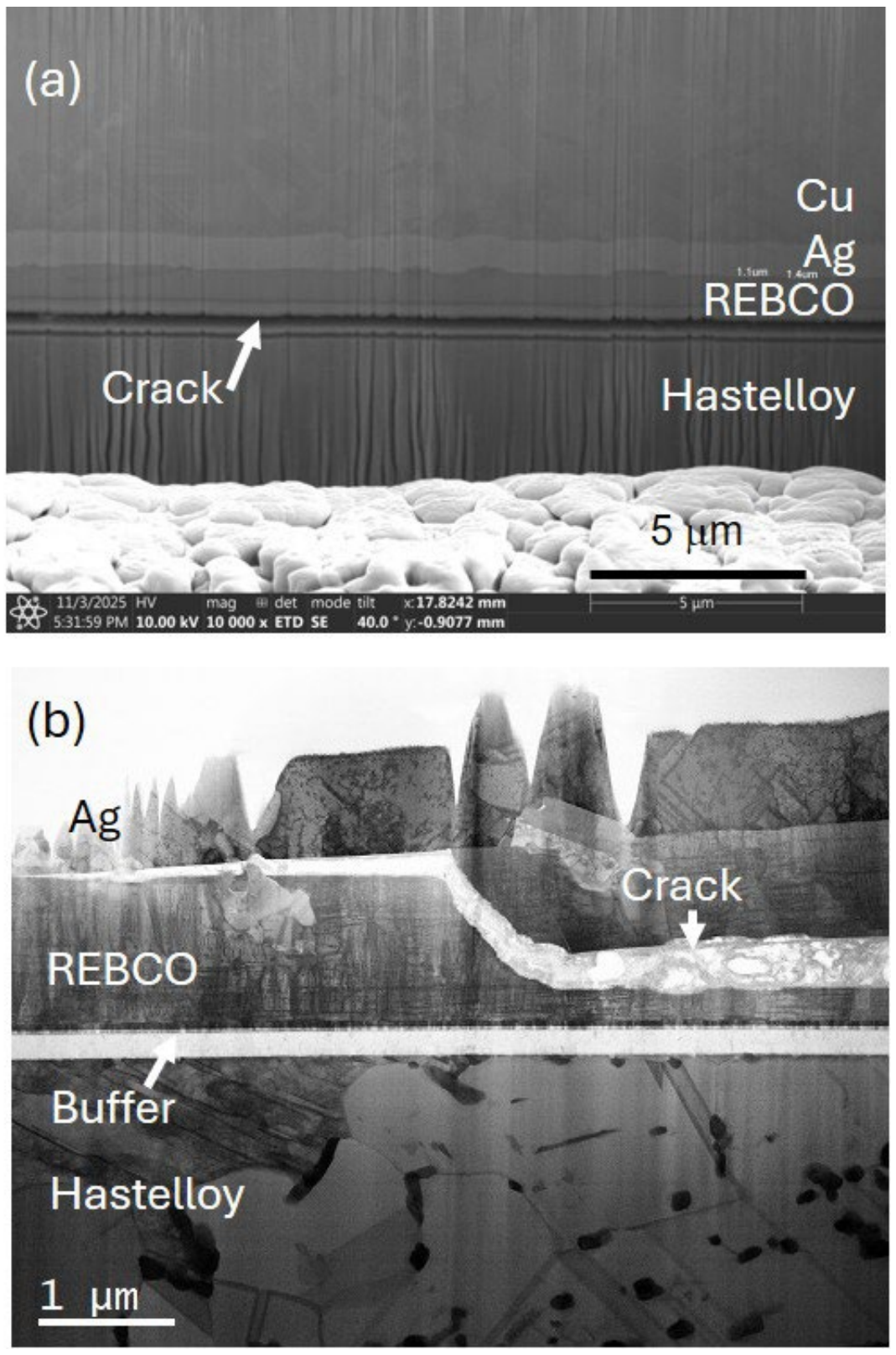

**Fig. 3.** (a) SEM image of the cross-section of a delaminated region. (b) TEM image of a region where delamination ends at silver/REBCO interface.

*B. Delamination in crossover joints of LSC1-2*

The degraded modules of LSC1-1 were replaced to construct LSC1-2 which was tested in 2026 and successfully reached 645 A. As the test campaign progressed, however, several crossover joints showed increasingly high resistance. This indicated REBCO degradation by electromagnetic stress at the joints. The stress at a crossover joint is explained by a simplified schematic in Fig. 4. During a magnetic field ramp, screening current is induced by the ramping radial field in the end modules. The stress by screening current makes the REBCO conductor slightly tilted as shown in the figure. The short piece of REBCO depicted in green is the 'bridge' piece between two modules, which will experience radial tensile and compressive stresses as indicated in the figure respectively. If the field ramp is reversed, the regions of tensile and compressive stress are switched. Evidently, the radial tensile stress may cause delamination on both sides of the crossover joint.

A postmortem examination of a crossover between module 21 and 22 was performed. The 'bridge', which consists of three 4 mm wide REBCO tapes (Fig. 5 (a)), was carefully desoldered on a 250 °C hot plate. After that, the residual solder was removed chemically by solder stripper 200 (Hubbard-Hall Inc.). The results are shown in Fig. 5. A crease can be seen in the middle of the 'bridge' (Fig. 5(a)) and surface bubbles are observed on both sides of the crossover joint (Fig. 5(b)).

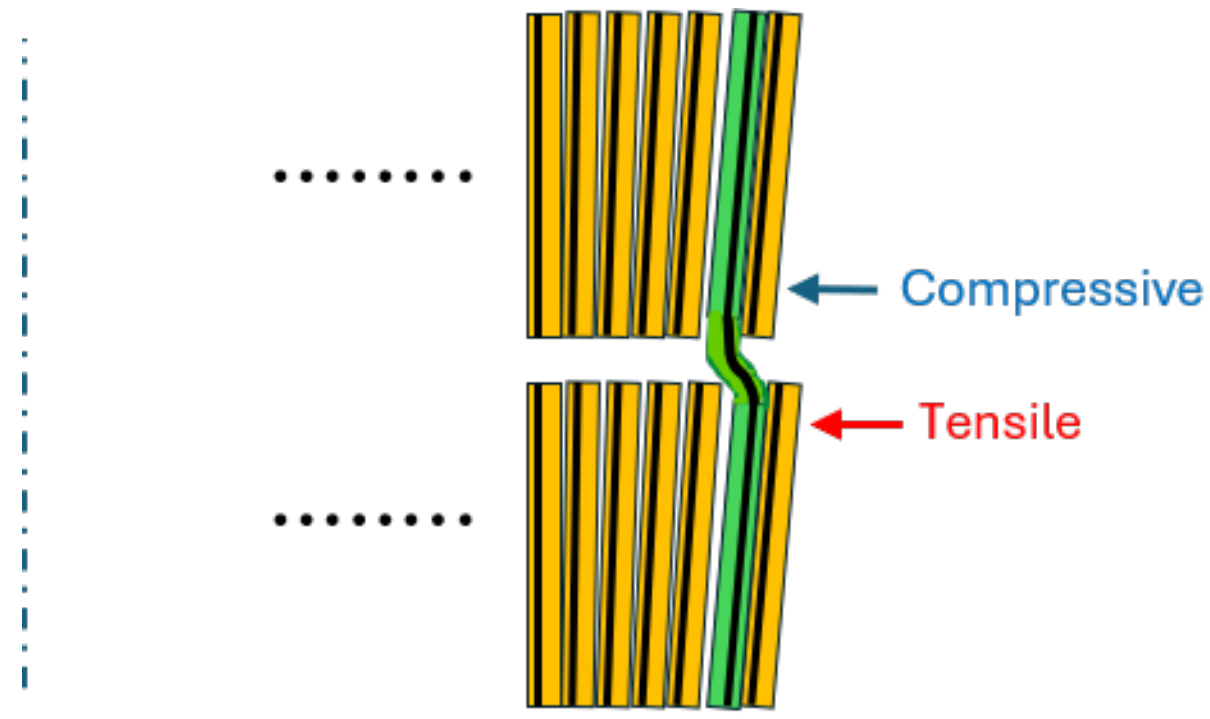

**Fig. 4.** Illustration of deformation of a crossover joint in end modules.

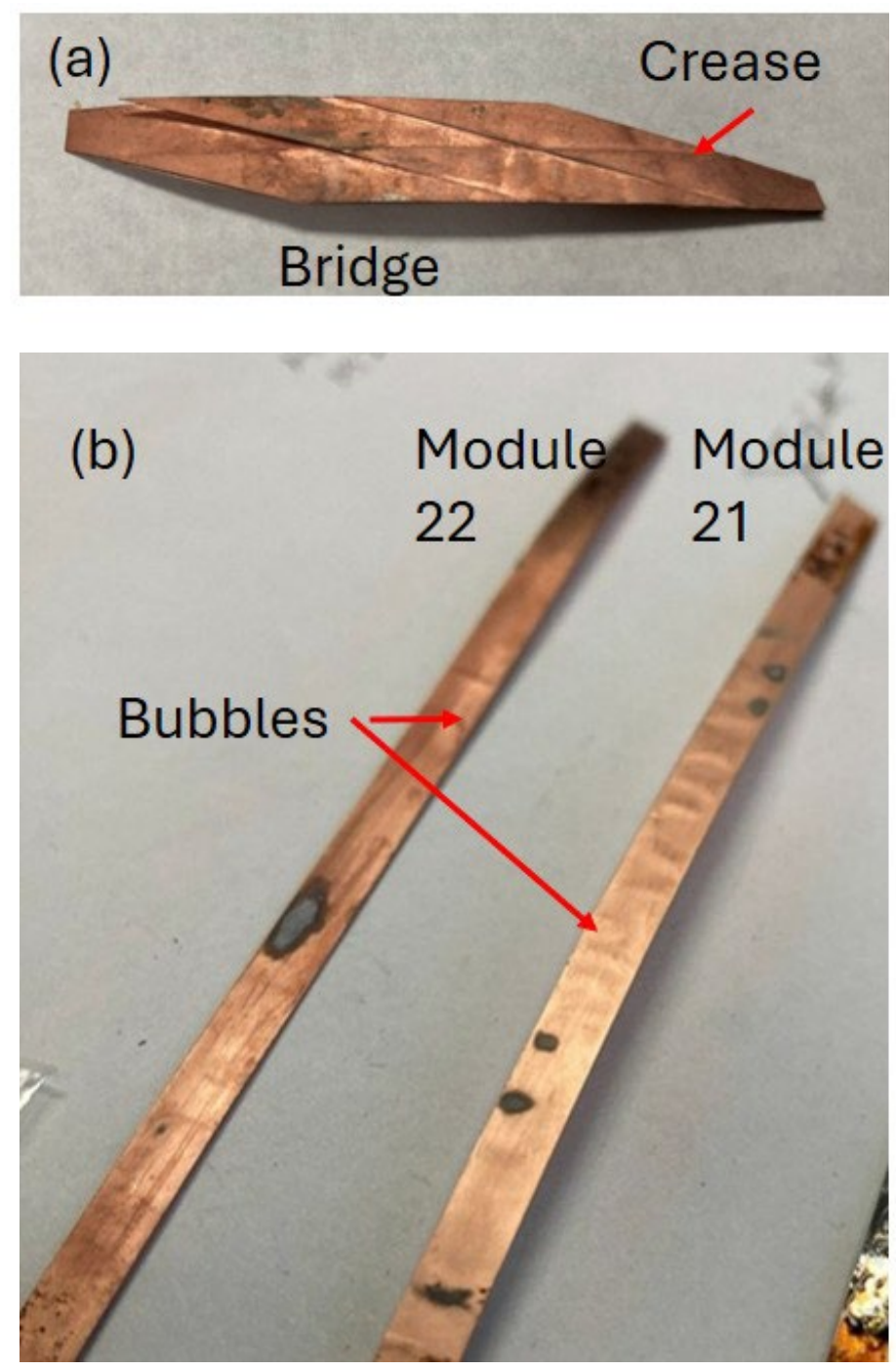

**Fig. 5.** (a) the joining piece and (b) base tape in the crossover joint regions both showing delamnation bubbles.

*C. Bubbles reproduced by torque magnetometry*

Two 'island' samples were tested by torque magnetometry, and delamination was first detected by sudden drop of torque values at about 2.5 MPa (in 7 T magnetic field) which is lower than previously measured 3.6 MPa [8]. The delaminated 'island' was completely separated from the rest of the sample and dropped to the bottom of the cryostat. Fig. 6 shows a picture of an 'island' after the test. For each sample, both sides of the delamination show black residual REBCO suggesting that delamination occurred mostly within the REBCO layer.

For the sample with surround edge copper, no delamination occurred up to 3.8 MPa (in 9 T magnetic field). At 12 T, the torque signal did not show a sudden drop but was much lower, indicating considerable degradation already occurred. After the test, a large bubble was observed on the sample surface as shown in Fig. 7 (a). When the copper layer was chemically removed, the bubble collapsed (Fig. 7(b)) indicating vacuum inside the bubble. The bubble was subsequently cut open to confirm the delamination (Fig. 7(c)). This torque magnetometry experiment proves that the delamination bubble can happen in REBCO by electromagnetic stress. Therefore, it is reasonable to attribute the bubbles found in LSC1-1 to electromagnetic stress.

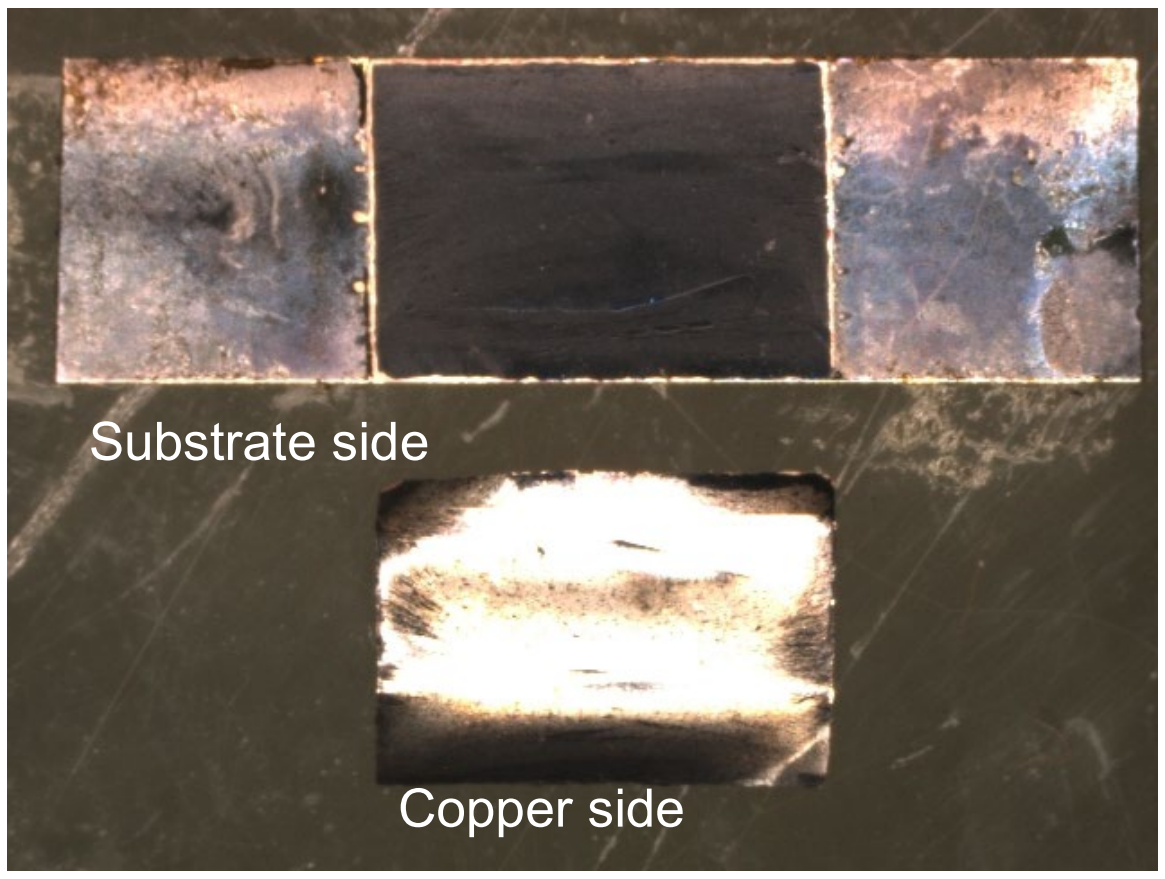

**Fig. 6.** A delaminated 'island' after a torque magnetometry test.

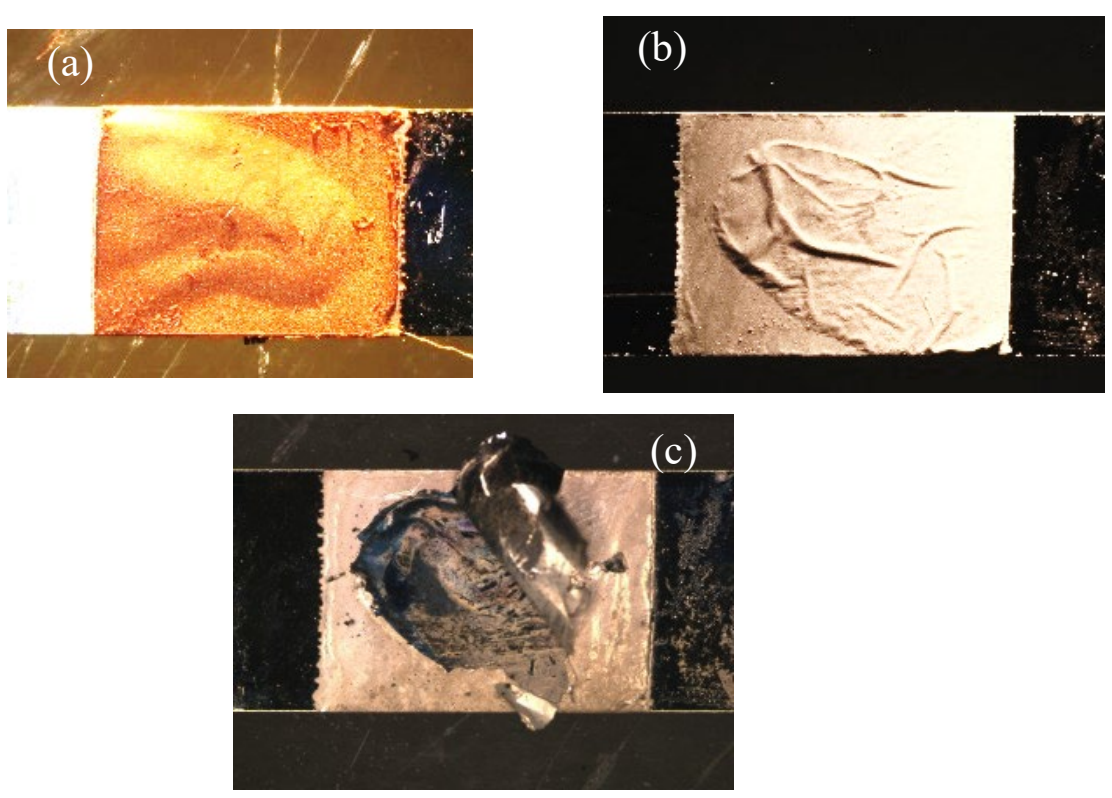

**Fig. 7.** A sample with surrounding edge copper after torque magnetometry test (a) bubble on conductor surface, (b) copper layer was removed, the bubble collapsed, (c) the bubble was cut open, delamination was confirmed.

## IV. Discussions

To find a solution to the delamination problem, we first need to understand the origin of delamination bubbles. Since the coil was wound with REBCO inside, the nominal transverse electromagnetic stress is compressive therefore cannot delaminate the conductor. The screening current, however, flows in both directions and results in both compressive and tensile transverse stresses on different areas of a conductor. Therefore, it is possible for modules where screening current is significant to be under tensile stress which may lead to delamination. In addition, during a magnet ramp or a magnet quench, the inductive voltage between the two REBCO tapes of the two-in-hand conductor can drive the negative current in one of REBCO tapes [12]. This can result in tensile stress in that tape.

To mitigate the delamination risk, it is desirable for the coil to be under sufficient radial pre-compression. This can be achieved by increasing winding tension for co-wind tapes and overband. For the delamination at the crossover joint, an improved design has been made for the next large scale test coil to reduce the displacement difference between modules at the critical areas indicated in Fig. 4.

Now we discuss stress distribution in the REBCO layer. Assuming a uniform current density through thickness, the accumulative electromagnetic stress linearly increases from zero at the REBCO/silver interface to maximum at the REBCO/buffer interface. The delamination strength presented here was the stress at the REBCO/buffer interface just before delamination occurred. If the delamination occurs within the REBCO layer, the stress at the initial delamination crack can be substantially lower. One can also speculate that the initial delamination crack occurred near the REBCO buffer interface where the accumulative stress is the highest. After the delamination, the pattern of screening current becomes fragmented and complicated, and the propagation of the initial crack becomes very difficult to predict. Further investigation into this subject is needed.

In our torque magnetometry test, the sample was only mechanically clamped at two ends where copper and REBCO layers were all removed (Fig. 1). During a test, the torque can slightly twist the sample. This twist significantly reduces its thermal contact with the sample stage, making the sample prone to flux jumps. These flux jumps were indicated by about 0.3 K temperature rise measured by a Cernox® sensor mounted on the sample stage. In a torque magnetometry data set, it is difficult to differentiate between a delamination event and a flux jump; both can damage the REBCO layer.

The torque magnetometer measured delamination strength to be 3.8 and 2.5 MPa for samples with and without surrounding edge copper respectively. This positive effect of surrounding edge copper on delamination agrees with the results in [13].

## V. Conclusion

Delamination bubbles were observed in insulated and paraffin impregnated large scale REBCO coils and crossover joints between adjacent double-pancake modules. They were caused by electromagnetic stress during coils tests at 4.2 K in an 11.4 T background field. A similar bubble was reproduced in a short REBCO sample by a torque magnetometry measurement in similar conditions. We found that the delamination occurs within the REBCO layer, not at the REBCO/buffer interface as one would expect.

## Acknowledgment

We thank Ernesto Bosque, Kurt Cantrell, Scott Marshall, Denis Markiewicz, Justin Deterding, Erick Aroyo, Mark Bird and others of the NHMFL 40 T project team for technical support in LSC coil design and testing. We also acknowledge SuperPower for postmortem inspection using Tapestar$^{TM}$.